\documentstyle[12pt,epsfig]{article}

\begin{document}

\baselineskip=6.8mm

%------------------------------------------------------------
\def\ap#1#2#3{           {\it Ann. Phys. (NY) }{\bf #1} (19#2) #3}
\def\arnps#1#2#3{        {\it Ann. Rev. Nucl. Part. Sci. }{\bf #1} (19#2) #3}
\def\cnpp#1#2#3{        {\it Comm. Nucl. Part. Phys. }{\bf #1} (19#2) #3}
\def\apj#1#2#3{          {\it Astrophys. J. }{\bf #1} (19#2) #3}
\def\asr#1#2#3{          {\it Astrophys. Space Rev. }{\bf #1} (19#2) #3}
\def\ass#1#2#3{          {\it Astrophys. Space Sci. }{\bf #1} (19#2) #3}

\def\apjl#1#2#3{         {\it Astrophys. J. Lett. }{\bf #1} (19#2) #3}
\def\ass#1#2#3{          {\it Astrophys. Space Sci. }{\bf #1} (19#2) #3}
\def\jel#1#2#3{         {\it Journal Europhys. Lett. }{\bf #1} (19#2) #3}

\def\ib#1#2#3{           {\it ibid. }{\bf #1} (19#2) #3}
\def\nat#1#2#3{          {\it Nature }{\bf #1} (19#2) #3}
\def\nps#1#2#3{          {\it Nucl. Phys. B (Proc. Suppl.) }
                         {\bf #1} (19#2) #3} 
\def\np#1#2#3{           {\it Nucl. Phys. }{\bf #1} (19#2) #3}
\def\pl#1#2#3{           {\it Phys. Lett. }{\bf #1} (19#2) #3}
\def\pr#1#2#3{           {\it Phys. Rev. }{\bf #1} (19#2) #3}
\def\prep#1#2#3{         {\it Phys. Rep. }{\bf #1} (19#2) #3}
\def\prl#1#2#3{          {\it Phys. Rev. Lett. }{\bf #1} (19#2) #3}
\def\pw#1#2#3{          {\it Particle World }{\bf #1} (19#2) #3}
\def\ptp#1#2#3{          {\it Prog. Theor. Phys. }{\bf #1} (19#2) #3}
\def\jppnp#1#2#3{         {\it J. Prog. Part. Nucl. Phys. }{\bf #1} (19#2) #3}

\def\rpp#1#2#3{         {\it Rep. on Prog. in Phys. }{\bf #1} (19#2) #3}
\def\ptps#1#2#3{         {\it Prog. Theor. Phys. Suppl. }{\bf #1} (19#2) #3}
\def\rmp#1#2#3{          {\it Rev. Mod. Phys. }{\bf #1} (19#2) #3}
\def\zp#1#2#3{           {\it Zeit. fur Physik }{\bf #1} (19#2) #3}
\def\fp#1#2#3{           {\it Fortschr. Phys. }{\bf #1} (19#2) #3}
\def\Zp#1#2#3{           {\it Z. Physik }{\bf #1} (19#2) #3}
\def\Sci#1#2#3{          {\it Science }{\bf #1} (19#2) #3}
\def\n.c.#1#2#3{         {\it Nuovo Cim. }{\bf #1} (19#2) #3}
\def\r.n.c.#1#2#3{       {\it Riv. del Nuovo Cim. }{\bf #1} (19#2) #3}
\def\sjnp#1#2#3{         {\it Sov. J. Nucl. Phys. }{\bf #1} (19#2) #3}
\def\yf#1#2#3{           {\it Yad. Fiz. }{\bf #1} (19#2) #3}
\def\zetf#1#2#3{         {\it Z. Eksp. Teor. Fiz. }{\bf #1} (19#2) #3}
\def\zetfpr#1#2#3{     {\it Z. Eksp. Teor. Fiz. Pisma. Red. }{\bf #1} (19#2) #3}
\def\jetp#1#2#3{         {\it JETP }{\bf #1} (19#2) #3}
\def\mpl#1#2#3{          {\it Mod. Phys. Lett. }{\bf #1} (19#2) #3}
\def\ufn#1#2#3{          {\it Usp. Fiz. Naut. }{\bf #1} (19#2) #3}
\def\sp#1#2#3{           {\it Sov. Phys.-Usp.}{\bf #1} (19#2) #3}
\def\ppnp#1#2#3{           {\it Prog. Part. Nucl. Phys. }{\bf #1} (19#2) #3}
\def\cnpp#1#2#3{           {\it Comm. Nucl. Part. Phys. }{\bf #1} (19#2) #3}
\def\ijmp#1#2#3{           {\it Int. J. Mod. Phys. }{\bf #1} (19#2) #3}
\def\ic#1#2#3{           {\it Investigaci\'on y Ciencia }{\bf #1} (19#2) #3}
\def\tp{these proceedings}
\def\pc{private communication}
\def\ip{in preparation}
\relax

\newcommand{\GeV}{\,{\rm GeV}}
\newcommand{\MeV}{\,{\rm MeV}}
\newcommand{\keV}{\,{\rm keV}}
\newcommand{\eV}{\,{\rm eV}}
\newcommand{\Tr}{{\rm Tr}\!}
\renewcommand{\arraystretch}{1.2}
\newcommand{\beq}{\begin{equation}}
\newcommand{\eeq}{\end{equation}}
\newcommand{\beqa}{\begin{eqnarray}}
\newcommand{\eeqa}{\end{eqnarray}}
\newcommand{\ba}{\begin{array}}
\newcommand{\ea}{\end{array}}
\newcommand{\bmat}{\left(\ba}
\newcommand{\emat}{\ea\right)}
\newcommand{\refs}[1]{(\ref{#1})}
\newcommand{\ler}{\stackrel{\scriptstyle <}{\scriptstyle\sim}}
\newcommand{\ger}{\stackrel{\scriptstyle >}{\scriptstyle\sim}}
\newcommand{\lag}{\langle}
\newcommand{\rag}{\rangle}
\newcommand{\ns}{\normalsize}
\newcommand{\cm}{{\cal M}}
\newcommand{\gr}{m_{3/2}}
\newcommand{\p}{\partial}

\def\rp{ $R_P$} 
\def\321{$SU(3)\times SU(2)\times U(1)$}
\def\tl{{\tilde{l}}}
\def\tL{{\tilde{L}}}
\def\bd{{\overline{d}}}
\def\tL{{\tilde{L}}}
\def\a{\alpha}
\def\b{\beta}
\def\g{\gamma}
\def\c{\chi}
\def\d{\delta}
\def\D{\Delta}
\def\db{{\overline{\delta}}}
\def\Db{{\overline{\Delta}}}
\def\e{\epsilon}
\def\l{\lambda}
\def\n{\nu}
\def\m{\mu}
\def\nt{{\tilde{\nu}}}
\def\p{\phi}
\def\P{\Phi}
\def\x{\xi}
\def\r{\rho}
\def\s{\sigma}
\def\t{\tau}
\def\th{\theta}
\def\ne{\nu_e}
\def\nm{\nu_{\mu}}
\def\rp{$R_P$}
\def\mp{$M_P$}     
\renewcommand{\Huge}{\Large}
\renewcommand{\LARGE}{\Large}
\renewcommand{\Large}{\large}
%\date{\today}
%\maketitle
%\vskip 2.0truecm

\begin{flushright}
hep-ph/9806376\\
PRL-TH-98/005
\end{flushright}
\vskip 2.0truecm
\begin{center}

{\Large \bf Rescuing Grand Unification Scenario for Neutrino Mass}

\vskip 2.0truecm

{ Anjan S.~Joshipura$^1$  and A. Yu. Smirnov$^{2, 3}$}\\
\vskip 0.5truecm

{\ns \it $^{1)}$ Inst. f\"ur  Theoretical Physik, Univ. of Dortmund,
44221, Dortmund, Germany\footnote{Alexander von Humboldt fellow, 
on leave from  Physical Research Laboratory, Ahmedabad,  India}}\\

\vskip 0.2truecm

{\ns \it $^{2)}$ Abdus Salam International Centre for Theoretical Physics,
I-34100 Trieste, Italy}\\

\vskip 0.2truecm

{\ns \it $^{3)}$ Inst. for Nucl. Research, Russian Academy of Sciences,
107370 Moscow, Russia}

\vspace{2cm}

\begin{abstract}
{\ns The see-saw mechanism for neutrino masses based on the 
Grand Unification   leads to the mass of the heaviest 
neutrino ($\approx \nu_{\tau}$) in
the range $(2 - 3)\cdot 10^{-3}$ eV and hence to a solution of the solar
neutrino problem through the $\nu_e-\nu_\t$ conversion. 
We suggest the existence of a light singlet fermion $S$ which 
mixes predominantly with muon neutrino through the mixing mass  
$m_{\mu s} \sim O(1)$ eV. The introduction of such a singlet 
allows one  (i) to solve the atmospheric neutrino problem 
via the $\nu_{\mu} \leftrightarrow S$ oscillations, (ii) to 
explain the LSND result and (iii) to get two component 
hot dark matter in the Universe. The phenomenology 
of this scenario is considered and  
the origin of the singlet as well as of the  scale
$m_{\m s}$ is discussed.} 
\end{abstract}

\end{center}

\thispagestyle{empty}

\newpage

%%%%%%%%%%%%%%%%%%%%%%%%%%%%%%%%%%%%%%%%%%%%%%%%%%%%%%%%%%%%%%%%%%%%%%%
\section{Introduction} 

Grand Unified Theories containing heavy right handed neutrinos 
provide a  natural framework for 
generation of small neutrino masses. 
Their generic properties are \cite{GU} 

(i) the (approximate) quark-lepton symmetry which relates
  the  quarks mass matrices and 
the Dirac neutrino mass matrix at the Grand Unification (GU) scale 
$\Lambda_{GU}$: 
$m^D_{\nu} \sim  m_q$. This relation could be exact in case of the third
generation so that $m^D_{3} =  m_t$ at $\Lambda_{GU}$; 

(ii) the lepton  number violation 
 at the Grand Unification scale. 

This naturally realizes   
the see-saw mechanism of the neutrino mass generation with  
the Majorana masses of the right handed neutrinos at the GU scale:  
$M_R \sim \Lambda_{GU} \sim 10^{16}$ GeV. 
In this scenario, the mass $m_3$ of the heaviest active neutrino 
 turns out to be in the range $\approx (1 - 3)\cdot 10^{-3}$ eV 
required for a solution of the solar neutrino problem 
via $\nu_e \rightarrow \nu_{\tau}$ resonance conversion \cite{MSW}.  
Moreover, the $\nu_e - \nu_{\tau}$ mixing  can also be 
\cite{BLS,SM1} in the correct range. 
 The parameters of the 
$\nu_e - \nu_{\mu}$ mixing in this scenario 
can be such that the solar neutrinos 
undergo also long range vacuum oscillations on the 
way to the Earth \cite{BLS,LP}. The interplay of conversion 
and oscillations  can lead to observable effects \cite{BLS}.  
We will refer to this possibility as  the Grand Unification 
scenario. 

The GU scenario does not allow one to explain 
the atmospheric neutrino problem. The existence of this problem and its 
oscillation solutions have been   reinforced recently by
high statistics observations of the zenith angle
asymmetry and $E/L$ dependence at the Super-Kamiokande \cite{SK} 
as well as by the SOUDAN data \cite{soudan}.   
Furthermore, in the GU scenario  
the neutrinos are  too light to explain the 
LSND \cite{LSND} result and to be a hot dark matter relevant 
for structure formation in the Universe.  

We will show in this paper  
that a minimal extension of the 
GU scenario obtained by introducing only one light 
singlet fermion which mixes  strongly with 
muon neutrino leads 
to solution of  all these problems simultaneously.  
In what follows we will describe  the scenario (sect. 2) 
and consider its phenomenology (sect. 3). In  sect. 4 
we will discuss the origin of singlet and its mixing.

%%%%%%%%%%%%%%%%%%%%%%%%%%%%%%%pp2%%%%%%%%%%%%%%%%%%%%%%%%%%%%%%%%%%%%%%%%

\section{Scenario}

In analogy with the quark sector let us start with 
basis $\nu_a = (\nu_{e'} , \nu_{\mu'},  \nu_{\tau'})$ 
where the
leptonic mass matrices  (both neutrino and charged lepton) have
hierarchical structure with small mixing. 
As in the GU scenario,  we suggest that active neutrinos 
acquire a mass term via the see-saw mechanism, 
\beq 
\label{sees} 
m^{S} = - m_D M_R^{-1} m_D^T~. 
\eeq 
Here $m_D$ and  $M_R$ are the Dirac and the Majorana mass matrices 
respectively.  

The scale in $M_R$ is identified with the grand unification scale
$\Lambda_{GU}$ and  the Dirac mass of the third generation is identified
with
the top quark mass. This leads to $ m_{\tau \tau}\sim {\mbox few}
\times 10^{-3}$ eV. In  case of the linear  hierarchy of the 
mass eigenstates of $M_R$, the second diagonal mass 
$m_{\mu \mu} \sim 10^{-5}$ eV  and  $m_{e e} \ll m_{\m\m}$.

We suggest that on the top of  the  GU  scenario  there 
exists an  additional light singlet state $\nu_s$  which couples  
predominantly to $\nu_{\mu'}$ in 
the basis $\nu_a$: 
\beq
\label{coupl}
m_{\mu s} {\nu}_{\mu'}^T \nu_s + m_{s s} \nu^T_s \nu_s + h.c. , 
\eeq
where 
\beq
m_{\mu s} \sim O({\rm  eV}), ~~~~m_{s s} \ll m_{\mu s}. 
\eeq  
The additional term $m_{\mu
\mu}{\nu}_{\mu'}^T{\nu}_{\mu'}$
can also be generated in some variants.  We will take $m_{\mu
\mu} \ll m_{\m s}$.   
The complete neutrino mass matrix can be parameterized as 
\beq
\label{4by4}
{\cal M}_\n = m^{S} + m(\nu_s) =  \left(
\begin{array}{cccc}
m_{ee}&m_{e\mu}&m_{e\t}&0\\
m_{e\mu}&m_{\m\m}&m_{\m \t}&m_{\m s}\\
m_{e\t}&m_{\m\t}&m_{\t\t}&0\\
0&m_{\m s}&0&m_{ss}\\ \end{array} \right )~.  
\eeq
In general, the elements 
$m_{e s}$ and $m_{\tau s}$ are  non-zero. In this case the matrix  
can be reduced to the above form by additional rotation of usual 
neutrino components. 
We assume that this rotation is not substantially bigger than rotation 
corresponding to usual (CKM-) mixing.  
The mass $m_{ss}$ can be generated if $\nu_s$ couples to the 
right handed neutrinos: $m_{sR} \nu_s \nu_R$. This 
leads to $m_{ss} \sim  m_{sR}^2/\Lambda_{GU}$
which gives $m_{ss} \sim 10^{-1}$ eV for $m_{sR} \sim 1$ TeV.

Let us introduce  $4\times 4$ matrix $U^{\nu}$ which diagonalizes 
(\ref{4by4}), so that $\nu_\a =  U^\n_{\a i} \n_i$, where 
$\n_i$ $(i = 1, 2, 3, 4)$ are the neutrino mass eigenstates.
Analogously, let $U^l$ be a $4\times 4$ matrix such that 
its $3\times 3$ block 
diagonalizes the charged
lepton mass matrix: $l_a = U^{l, 3}_{a \a} l_{\a}$,  
($l_{\a} = e, \mu, \tau$) and 
$U^{l}_{\a a} = \delta_{\alpha s} \delta_{a s}$. 
Then the leptonic CKM   matrix  
determining the flavor mixing of neutrinos: 
$\nu_{\a} = U_{\a i}\nu_i$ equals 
\beq 
\label{mixing}
U_{\a i}=U^{*l}_{\a a} U^\n_{a i}, ~~~\a = e, \mu, \tau, s ~~~i = 1, 2, 3,
4; ~~~~a = e', \mu', \tau', s ,    
\eeq

The introduction of  $\nu_s$ with couplings 
(\ref{coupl}) substantially changes the neutrino mass spectrum and 
lepton mixing of the  GU scenario.

(i) Since $m_{\mu s}$ is much bigger than all see-saw contributions, 
$\nu_{\mu'}$ and $\nu_s$ will form pseudo Dirac neutrino 
with mass $\approx m_{\mu s}$. The states  $\nu_{\mu'}$ and $\nu_s$ 
mix maximally in $\nu_2$ and $\nu_4$ and the mass squared
difference  equals 
\beq
\label{ma}
\Delta m^2_{24} \equiv m_4^2 - m_2^2  = 2 m_{\mu s} (m_{\m\m}+m_{ss}).  
\eeq
We fix $\Delta m^2_{24} = (10^{-3} - 10^{-2})~ {\rm eV}^2$,  so that
oscillations $\nu_{\mu} \leftrightarrow \nu_s$ 
solve the atmospheric neutrino problem \cite{SK}. For $m_{\mu s}\sim \eV$,
this leads to  $(m_{\m\m} + m_{ss}) \sim (0.005-0.05)$ eV. 

(ii) The introduction of $\nu_s$ does not  practically change the 
mass and mixing of the $\nu_{e'} - \nu_{\tau'}$ 
sub-system. Indeed, diagonalization of the 
$\nu_\m-\nu_s$ sub-system will give corrections to the mass terms of the 
$\nu_{e'}$ and  $\nu_{\tau'}$ suppressed by 
$m_{e \mu}/m_{\mu s} \ll 10^{-3}$ and 
$m_{\mu \tau}/m_{\mu s} \ll 1$. Hence, 
$U^\n_{a i} \approx U^{ss}_{a i}$ for $a = e',  \tau', i = 1,  3$, where
$U^{ss}$ diagonalizes the purely see-saw contribution, eq. (\ref{sees}).
Therefore as in the GU scenario \cite{BLS} the $\nu_{e}-\nu_{\t}$
conversion solves the solar neutrino problem \cite{BLS}.

(iii) Unlike the $\nu_{e'}-\nu_{\t'}$ sub-system, the mixing between
$\nu_{e'}-\nu_{\m'}$ as well as $\nu_{\mu'} - \nu_{\t'}$
following from 
(\ref{4by4}) is  suppressed in comparison with the see-saw
prediction by factors : $m_{\mu \mu}/m_{\mu s} \ll 10^{-3}$ and
$m_{\tau \tau}/m_{\mu s} \ll 10^{-3}$ respectively. 
Therefore, the contribution of the neutrino mass matrix to lepton mixing 
in $e - \mu$ and $\mu - \tau$ sectors is negligibly small. The mixing will
be
determined mainly by the mass matrix of the charged leptons, 
$$
\theta_{e \mu} \approx \theta_{e \mu}^l ~~~~~
\theta_{\mu \tau} \approx \theta_{\mu \tau}^l .  
$$
$\theta_{e \mu}^l$ can be chosen to explain the 
LSND result. 

Thus, we can write the  matrix $U^\n_{a i}$ which 
diagonalizes (\ref{4by4}) and therefore 
describes  
the contribution of the neutrino mass matrix to the lepton mixing 
as 
\beq
\label{44nu}
U^\n \approx 
\left(
\begin{array}{cccc}
\cos \theta  & 0 & \sin\theta & 0  \\
0 & \frac{1}{\sqrt 2} & 0 & \frac{1}{\sqrt 2}\\
- \sin\theta  & 0 &  \cos \theta  & 0\\
0 & - \frac{1}{\sqrt 2} & 0 & \frac{1}{\sqrt 2}\\  
\end{array} \right )~, 
\eeq 
where $\sin \theta\  = U_{e'3}^{\nu}$ 
describes the $\nu_{e'}- \nu_{\tau'}$  mixing. 
 For the charged lepton matrix $||U^{l}_{\a a}||$ we assume 
 a form typical for the quark mixing. Then the total lepton mixing 
matrix  defined in (\ref{mixing}) can be written as  
\beq
\label{mixing1}
U \approx
\left(
\begin{array}{cccc}
U_{e e'}^* c  - U_{e \tau'}^* s & 
U_{e \mu'}^*/\sqrt{2} & 
U_{e e'}^* s  + U_{e \tau'}^* c & 
U_{e \mu'}^*/\sqrt{2} \\
U_{\mu e'}^* c  - U_{\mu \tau'}^* s & 
U_{\mu \mu'}^*/\sqrt{2} & 
U_{\mu e'}^* s  + U_{\mu \tau'}^* c & 
U_{\mu \mu'}^*/\sqrt{2} \\
U_{\tau e'}^* c  - U_{\tau \tau'}^* s & 
U_{\tau \mu'}^*/\sqrt{2} & 
U_{\tau e'}^* s  + U_{\tau \tau'}^* c & 
U_{\tau \mu'}^*/\sqrt{2} \\
0 & - 1/\sqrt 2 & 0  & 1/\sqrt 2\\
\end{array} \right )~. 
\label{matrix}
\eeq
Here $c \equiv \cos \theta, s \equiv \sin \theta$. 
Notice that in our approximation $\nu_s$ has no admixture in the light
neutrino states $\nu_1$ and $\nu_3$. Moreover, 
\beq
\label{equality}
U_{\alpha 2} =  U_{\alpha 4} , ~~~ \alpha = e, \mu, \tau . 
\label{u2u4}
\eeq

The resulting mass spectrum and mixing pattern are shown in Fig. 1.
It is seen that this mass spectrum 
simultaneously solves the atmospheric neutrino problem,   
explains the LSND result and provides two component HDM in the Universe.

%%%%%%%%%%%%%%%%%%%%%%%%%%%%%%%%pp3%%%%%%%%%%%%%%%%%%%%%%%%%%%%%%%%%%%%%

\section{Phenomenology}

For generic values of parameters the suggested 4 neutrino scheme has
three different scales of the mass squared differences with the following
hierarchy: 
$$
\Delta m^2_{12}\gg  
\Delta m^2_{24}\gg 
\Delta m^2_{13} . 
$$
Moreover, one mixing is almost maximal, whereas 
other mixings are small. This simplifies the task reducing 
dynamics in all particular cases to dynamics of two 
neutrino systems. Phenomenology of this scheme 
has been considers partly in \cite{hier} also.

%%%%%%%%%%%%%%%%%%%%%%%%%%%%%%%%solarnu%%%%%%%%%%%%%%%%%%%%%%%%%%%%%%%%%%%

\subsection{Solar neutrinos}

For solar neutrinos, the heavy components $\nu_2$ and $\nu_4$ 
``decouple" leading to the averaged oscillation effect, 
while the combination of the light states 
\beq
\nu_e' \equiv {1\over \cos \phi_e}  (U_{e1}\nu_1+U_{e3}\nu_3) 
\eeq
with  $\cos^2 \phi_e \equiv U_{e1}^2+U_{e3}^2$ gets converted to 
its  orthogonal state. 
The survival probability can be readily written as 
\beq
\label{solarp}
P_{\nu_e} = \cos^4\phi_e P_{2} + \frac{1}{2} \sin^4 \phi_e ~, 
\eeq 
where $P_{2}$ is the two generation survival  probability 
for $\nu_e' \rightarrow$(orthogonal state);   
here we have used  
 $U_{e2}^2 =  U_{e4}^2 = 
0.5 \sin^2 \phi_e$ following from eq.(\ref{u2u4}). Strong bound on the
angle $\phi_e$ 
follows from the short range laboratory experiments.  
Indeed, for the short range experiments we find the 
survival probability 
\beq
\label{shot1}
P_{\nu_e} = 1  -  \sin^2 2\phi_e \sin^2 \frac{\Delta m_{14}^2 t}{4 E}.  
\eeq
{}From BUGEY bound \cite{BUGEY} $\sin^2 2\phi_e \leq (4 - 8) \cdot
10^{-2}$
for 
$\Delta m_{14}^2 \sim O( {\rm eV})$ we get 
$\cos^2 \phi_e \geq 0.98$  and 
$0.5 \sin^4 \phi_e \leq (0.5 - 2) \cdot 10^{-4}$.
Thus to a good approximation 
$P_{\nu_e} = P_2 (\Delta m^2_{13}, \sin^2 2\theta_e)$, where 
$\sin^2 2\theta_e \equiv 4U_{e1}^2 U_{e3}^2 /\cos^2 \phi_e$.  
The corrections due to the presence of heavy states are very small. 
The electron neutrino survival 
probability does not significantly
differ from the corresponding two active neutrino case  
which requires 
\beq
\label{solar}
\Delta m_{13}^2\sim (4 - 10) \times 10^{-6} \eV^2\;
\; ; \;\; 4 U_{e1}^2 U_{e3}^2  \sim (3-10) \times 10^{-3}. 
\eeq

%%%%%%%%%%%%%%%%%%%%%%%%%%%%%%%atmosnu%%%%%%%%%%%%%%%%%%%%%%%%%%%%%%%%%%

\subsection{Atmospheric neutrinos}

The atmospheric neutrino deficit is explained mainly by the 
oscillations $\nu_{\mu}\leftrightarrow \nu_{s}$. 
The transition probabilities are small in  other channels of oscillations.
As follows from eq.(\ref{mixing1}), the muon neutrino 
has a small admixture of the lighter states 
$\nu_1$ and $\nu_3$. This leads to  oscillations 
due to large splitting $\Delta m_{12}^2$ which get averaged out.
Strong non averaged oscillations are stipulated 
by splitting $\Delta m_{24}^2$ while the splitting $\Delta m_{31}^2$ is 
irrelevant due to smallness 
of the $\nu_e$ mixing in the heaviest states. As a result, we find
the survival probability for the muon neutrinos: 
\beq 
\label{ats}
P_{\nu_\mu} = 1 - \cos^4 \phi_{\mu} 
\sin^2{\Delta m_{42}^2t\over 4 E} - {1\over 2}\sin^2 2\phi_{\mu}~, 
\eeq
where $\cos^2 \phi_{\mu} \equiv  U_{\m 2}^2 + U_{\m 4}^2$,  
and we  used equality  $U_{\m 2}^2 =  U_{\m 4}^2$. 
The last term in (\ref{ats}) is due to averaged oscillation 
contribution. 
The modifications of the usual two neutrino oscillation probability 
in (\ref{atm})  described by the parameter $\phi_{\mu}$ 
are strongly constrained by the short-range experiments. 
Indeed, for the the oscillations with $\Delta m_{4i}^2\sim
\Delta m_{2i}^2\sim \eV^2 \;\;(i=1,3)$  
we find the survival probability 
\beq
\label{short}
P_{\nu_\mu} =  1 -   \sin^2 2\phi_{\mu} 
\sin^2{\Delta m_{41}^2t\over 4 E}. 
\eeq
The negative results in 
the $\nu_\m$ disappearance experiment  by CDHS group imply 
\cite{CDHS} 
$ \sin^2 2\phi_{\mu} <0.1 $ 
for $1.0 \eV^2<\Delta m_{41}^2<10 \eV^2$. 
Thus the last term in (\ref{ats}) is less than 0.05 and
$P_{\nu_\mu}$ reduces approximately  to two neutrino
case. 
In particular, 
the  double ratio  in  the atmospheric
neutrino fluxes  is determined by the  
survival probability $P_{\nu_\m}$. 
The analysis of the data \cite{FVY} gives 
relevant ranges of the neutrino parameters
\beq
\label{atm}
\Delta m_{42}^2\sim (2-10) \times 10^{-3} \eV^2\;\; ;\;\;
  4 U_{\m 2}^2U_{\m 4}^2
\sim (0.8-1)
\eeq 
which are similar to those for $\nu_{\mu} - \nu_{\tau}$. 

The $\nu_{\mu} - \nu_{\tau}$ 
and $\nu_{\mu} - \nu_s$ oscillation solutions are rather similar as
far as the low energy charged current data are concerned.  
There are  crucial points which distinguish the solutions. 

1. The neutral current
induced events are unaffected if the atmospheric
neutrinos oscillate to $\nu_\t$ but  
are reduced in  the case of the $\nu_{\mu} - \nu_s$ oscillations 
\cite{VS}. The neutral current reactions,  
in particular  
$\nu N \rightarrow \nu N \pi^0$,  give main 
contribution to the sample of isolated $\pi^0$ events. 
Useful measure of this
is provided by the ratios $(i) N_{\pi^0}/N_e \;(ii) N_{\pi^0}/N_\m$ 
and $(iii)
N_{\pi^0}/N_{multi-ring}$. Here $N_{e,\m,\pi^0}$ denote the standard
e-like, $\mu$-like events 
and number of events with isolated  pions produced mainly 
through neutral current interaction. The $N_{multi-ring}$ 
result mainly through  the charged
current induced reactions \cite{VS}. The  ratios can differ
by (30-50)\% in two relevant cases if the survival probability is
around 0.6 \cite{VS}. It should be possible to see this difference and
distinguish these two possibilities with increased statistics
in the  Superkamiokande experiment.

2. Events induced by the high energy neutrinos, like 
upward going muons (throughgoing and stopping) 
are affected by matter in the case of $\nu_{\mu} - \nu_s$ oscillations
and there is no matter effect in 
$\nu_{\mu} - \nu_{\tau}$ channel. 
For   $\nu_{\mu} - \nu_s$ oscillations the matter effect 
leads to peculiar zenith angle ($\Theta$) distribution of the muons  
\cite{LS,LMS}. The distribution has two dips at 
$\cos \Theta = - 0.8 - 0.3$ and $\cos \Theta = - 1.0 - 0.8$. 
The latter dip is due to the parametric resonance \cite{param} in
oscillations 
of neutrinos which cross the core of the Earth. In fact, 
MACRO, Baksan and SuperKamiokande data give some indications 
of such a distribution \cite{LMS}.  

3. Up-down asymmetry in neutral current induced events, i.e.
$\nu N \rightarrow \nu N \pi^0$ can also provide a means to distinguish
the two oscillation channels \cite{pak}. 
%%%%%%%%%%%%%%%%%%%%%%%%%%%%%lsnd%%%%%%%%%%%%%%%%%%%%%%%%%%%%%%%

\subsection{Short range $\nu_{\mu} - \nu_e$ and $\nu_{\mu} -
\nu_{\tau}$ oscillations} 

The transition probability $P_{\nu_e\nu_\m}$ 
in the short range experiments which are sensitive only to the large 
mass splitting  $\Delta m_{41}^2\approx \Delta m_{43}^2
\approx \Delta m_{23}^2\approx \Delta m_{21}^2$  
is given by
$$
P_{\nu_e\n_\m}= \sin^2 2\phi_{e \mu} \sin^2{\Delta m_{41}^2t\over 4 E},  
$$
where  according to (\ref{matrix})
$$
\sin^2 2\phi_{e \mu} = - 4(U_{e1} U_{\mu1} + U_{e3} U_{\mu3})
(U_{e2} U_{\mu2} + U_{e4} U_{\mu4}) \approx 4 (U_{e \mu'}^l U_{\mu
\mu'}^l)^2 . 
$$
Positive signal reported by LSND implies  
\begin{equation}
\label{lsnd}
\sin^2 2\phi_{e \mu} \sim (0.1 - 4)\cdot  10^{-2}, ~~~~ 
\Delta m_{41}^2 \sim (0.3 - 2.5)~ {\rm eV}^2.   
\end{equation}
Notice that
$ \theta_{e\m}\sim \theta^l_{e\m} \sim \sqrt{m_e/ m_\m}$
gives for the effective mixing angle appearing in (\ref{lsnd})
$$ 
\sin^2 2\phi_{e \mu}\sim 2\times  10^{-2}
$$
which is in the right range.   

For $\nu_{\mu} - \nu_\tau$ short range oscillations  
we get similar two neutrino oscillation probability with 
$\Delta m^2 = \Delta m^2_{14}$ and 
$$
\sin^2 2\phi_{\mu \tau} = - 4(U_{\tau 1} U_{\mu1} + U_{\tau 3} U_{\mu3})
(U_{\tau2} U_{\mu2} + U_{\tau4} U_{\mu4}) \approx 4 (U_{\tau \mu'}^l
U_{\mu \mu'}^l)^2 .
$$
The  depth of oscillations is determined basically by 
mixing of charged leptons in the selected basis.  

%%%%%%%%%%%%%%%%%%%%%%%%%%%%%%%%%%%%%%%%%%%%%%%%%%%%%%%%%%%%%%%%%%%%%%

\subsection {Supernova neutrinos}

Let us first find  the level crossing scheme, i.e.   
the dependence of the eigenvalues of the Hamiltonian 
of the system on the density of the medium. The Hamiltonian 
can be written as  
$$
H = \frac{M^2}{2 E} + V,  
$$
where $M$ is the neutrino mass matrix in flavor basis and $V = diag(V_e,
V_{\mu}, V_{\mu}, 0)$ is the matrix of the potentials. 

The levels can be found in the following way. 
One diagonalizes first the strongly mixed heavy sub-system 
$\nu_{\mu}$ and $\nu_s$. This gives the levels 
$\nu_{2m}'$, $\nu_{4m}'$. Then using smallness of all other mixings 
one gets the level crossings: 
$\nu_e - \nu_{4m}'$,  $\nu_e - \nu_{2m}'$,  
at large densities and $\nu_e -  \nu_{\tau}$ crossing at small 
density. There is no level crossing in the anti neutrino channels. 

In  case of the complete adiabaticity one predicts  
the following transitions: 
\begin{eqnarray} 
\nu_e \rightarrow \nu_{4m}' \approx (\nu_{\mu} + \nu_s)/\sqrt{2},\\ 
\nu_{\mu} \rightarrow \nu_e \rightarrow \nu_{\tau},\\ 
\nu_{\tau}  \rightarrow \nu_e.  
\end{eqnarray}
Thus at the {\it surface}  of the star one will have hard 
(corresponding to the original $\nu_{\tau}$ spectrum) 
electron neutrinos, $\nu_{\mu}'$s with soft spectrum 
and the flux of sterile neutrinos which is 
about 1/12 of the total  neutrino flux. No changes in the 
antineutrino channels are expected.     

The production of the heavy elements  due to $r$ - processes in the
supernovae imply that transitions 
$\nu_{\tau} \rightarrow \nu_e$,  $\nu_{\mu} \rightarrow \nu_e$
are suppressed in the inner parts of the star
\cite{qian}. As follows from the 
level crossing scheme  the appearance of $\nu_e$ can be 
due to adiabatic transition  $\nu_{\mu} \rightarrow \nu_e$. 
The  transition occurs above the region of the 
$r$-processes if $\Delta m^2 \approx m_{\mu s}^2 < 2$ eV$^2$. 
For larger mass square differences mixing angles of active neutrinos   
should be strongly suppressed.

\subsection{Primordial Nucleosynthesis} 

The oscillations $\nu_{\mu} \rightarrow \nu_s$ and  
$\nu_{\tau} \rightarrow \nu_s$ lead to 
formation of the equilibrium concentration of the 
sterile neutrinos in the Early Universe. 
Therefore by the time of the 
primordial nucleosynthesis there were $N_{\nu} = 4$ 
neutrino species. This is inconsistent with the bound 
$N_\n\leq 2.6$ based on low values of deuterium abundance \cite{low}
but is allowed by recent conservative estimates $N_{\nu} <4.5$ \cite{high}.   
The appearance  of  sterile neutrinos can be suppressed 
if there is sufficiently large leptonic asymmetry in 
the  Early Universe at the epoch with $T < 10 - 20 $ MeV . 
However, this asymmetry cannot be produced 
in   $\nu_{\tau} \rightarrow \nu_s$,  ${\bar\nu}_{\tau} \rightarrow
{\bar\nu}_s$ oscillations by mechanism suggested in \cite{FV} 
because of the negative sign of the mass squared difference 
$m_3^2 - m_2^2$ in the scenario under consideration.

%%%%%%%%%%%%%%%%%%%%%%%%%%%%%%%%%%%%%%%%%%%%%%%%%%%%%%%%%%%%%%%%%%%%%%%

\section{Origin of scales}

Both the origin and lightness of sterile state can be understood
in supersymmetric schemes \cite{qgf1,qgf2}. The sterile neutrino  may be
identified  with fermionic component of the Goldstone multiplet arising
due to  spontaneously broken global symmetry \cite{qgf1} such
as Peccei Quinn (PQ) symmetry or with some hidden sector field
\cite{qgf2} in string based models. Violation of the $R$
symmetry  in these theories  could lead to coupling of
the sterile  state to ordinary neutrino. 
The general discussion of the mass and couplings of such sterile state was
given in \cite{qgf1,qgf2}. Here we discuss a specific example which can
reproduce the mass and mixing pattern discussed above.

The example is based on the supersymmetric standard model with additional
$U(1)_Q$ symmetry. We need to add singlet fields $\s,\s'$ and $y$ in order
to implement the breaking of the additional symmetry. 
The  $U(1)_Q$ charge is defined as
$$ Q=B-L-X\;.$$ 
where $B$ and $L$ are the baryon and the lepton charges correspondingly. 
Keeping ultimate unification in mind, the $X$ charge is assumed 
to commute with $SO(10)$. The X charge can be normalized such that 
the common charge of the standard fermions is one unit.
The $X$-charges of the Higgs fields $(H_1,H_2,\sigma,\sigma',y)$ are then
chosen as $(-2,-2,2,-2,0)$. Note that the $U(1)_{Q}$ symmetry may be
identified
with the PQ symmetry. In addition to the standard Yukawa interactions, the  
following terms are allowed in the superpotential by the $U(1)_Q$. 
\beq
\label{sb}
W=\lambda y (\s\s'-f^2)+M_iN_iN_i +\e_iL_iH_2+\delta H_1H_2{\s^2\over
M_P}\;.
\eeq 
where $f$ and $\epsilon_i$ are the mass parameters, $M_P$ is the 
Planck mass,   and $\delta = O(1)$.  
The first term leads to the VEV  
$\langle \sigma \rangle = \langle \sigma' \rangle = f$   and thus  the
$U(1)_Q$ breaking which generates the Goldstone multiplet 
$S={(\s-\s')/ \sqrt{2}}$. The mass terms for the right
handed neutrinos simultaneously break $B-L$ and $X$ but preserve
$U(1)_Q$. It is natural to identify the masses $M_i$ with the GU scale in
the present context. The parameters $\e_i$ allowed by the $U(1)_Q$ 
automatically break  the $R$-parity. 
The last term generates the $\mu$ parameter of the
required magnitude:  $\mu = \delta f^2/ M_P$ 
and also leads to coupling of $S$ to neutrinos because
of the presence of  $\e_i$. We need to assume 
$\e_{1,3}\ll \e_2$ in order to ensure that $S$ predominantly couple to 
$\nu_\m$. 

It is easily seen  that the last two terms generate 
mixing of $\nu$, $S$ and Higgsinos. The block diagonalization of the
mixing matrix  gives \cite{qgf1}
\beq
\label{mus}
m_{\mu s}={c\e_\m v_2\over f}\; ,
\eeq
when  heavy Higgsino states are decoupled. Here $c\sim O(1)$. 

The  $\e_2$-term  also generates mixing between $\nu_\m$
and gaugino. This mixing in turn gives the contribution to 
the mass of the muon neutrino \cite{gauge},  
$m_{\m\m}$:
\beq
\label{mumu}
m_{\m\m} = 
\left( {\e_2\over \m}\right )^2 m_0
\eeq
The above  contribution dominates over the standard see-saw contribution
to $m_{\m\m}$
if the right handed neutrinos have masses at the GUT scale.  
The mass $m_0$ is completely determined by the parameters of  
the supersymmetric standard
model if one starts with the universal scalar masses at the GUT 
scale. $m_0$ is mainly determined in this case by the $b$ Yukawa
coupling $h_b$ and is typically given by
$$ 
m_0\sim {M_W^2\over m_{SUSY}}\left({3 h_b^2\over 16 \pi^2}
ln {M_{GUT}^2\over M_Z^2}\right)^2\; .
$$
Detailed analysis shows \cite{gauge} that 
$
m_0\sim \keV- \MeV
$,  
when MSSM parameters are varied over its allowed range. 

The breaking of supersymmetry results in generation of the direct mass
term
$m_{ss}$ for $S$. The scale of $m_{ss}$ need not coincide with the susy
breaking scale but could be much smaller, e. g.  ${m_{SUSY}^2/M_P}$
\cite{qgf2} or ${m_{SUSY}^2/M_{GUT}}$ depending upon the details of
underlying theory. The $m_{ss}$ is found \cite{qgf1,qgf2} to be suppressed
if 
({\em i}) all the MSSM fields and some or all 
of the singlet fields have no scale
kinetic terms  or ({\em ii}) if supersymmetry  breaking is 
mediated by the gauge interactions and the MSSM fields as well as
the singlets  $\s,\s',y$ and the right handed neutrinos do not
directly couple to the sector in which supersymmetry is broken.
({\em iii}) Singlet $S$ is identified with some of the modulie fields.
The $m_{ss}$ can arise radiatively in some cases \cite{qgf1} but its
magnitude can be 
smaller than the neutrino-gaugino mixing contribution to $m_{\m\m}$.
The latter contribution then sets the scale for the atmospheric neutrino
anomaly. 

The mixing mass 
$m_{\m s}$ is in the eV range if
$
{\e_2 / f}\sim 10^{-11}
$.
The $m_{\m\m}$ can be $\sim 10^{-2} \eV$ provided
\beq
\label{r2}
({\e_2 / \m})^2\sim 10^{-5}-10^{-8} . 
\eeq
It follows then that 
\beq
\label{values}
\e_2\sim (10^{-2} - 1  ) \GeV \;\;\; ; \; \; \; f\sim
 (10^9 -  10^{11}) \GeV . 
\eeq
The value of  $f$ required here is included in the allowed range for
the PQ symmetry breaking scale. 
The  hierarchy $\e_{1,3}\ll \e_2$ among the $R$ breaking
parameters can be explained by introducing  the additional 
(approximate) horizontal symmetry prescribing non-zero charges 
for the first and the third generation and zero charge for 
second generation and singlets. 

We have identified the singlet state with a supersymmetric partner of 
a Goldstone boson in the above example. In a more general context, singlet
may remain light for other reasons. The desired
value of the mixing mass  can be obtained in such cases 
from the following relation: 
$$
m_{\nu s} = 
\frac{<X>_F v}{\Lambda_{GU}^2} = 
\frac{m_{3/2} M_P v}{\Lambda_{GU}^2}, 
$$
where $<X>_F$ is the VEV of the field which breaks SUSY. 
This possibility can be realized if e.g. the term 
\beq
\frac{LSX^*H_d^*}{\Lambda_{GU}^2}
\label{term}
\eeq
exists in the Kahler potential (it might be if 
unification with gravity occurs at the Grand Unification scale). 
The effective interaction   (\ref{term}) can be generated 
due to exchange of the heavy superfield doublet $\Phi_u$ 
which has the  couplings  in the
superpotential:  $L S\Phi_u$ and  $X H_d  \Phi_u$.

%%%%%%%%%%%%%%%%%%%%%%%%%%%%%%%%%%%%%%%%%%%%%%%%%%%%%%%%%%%%%%%%%%%%%%

\section{Discussion and Conclusions}

The phenomenology of our  scenario strongly differs from the 
phenomenology of the 4-neutrino schemes 
suggested in  
 \cite{others,qgf1,qgf2}. There  solar neutrinos are converted into
sterile neutrinos, 
whereas the atmospheric neutrinos undergo 
$\nu_{\mu} - \nu_{\tau}$- oscillations. 
In contrast, in the present scheme  the
solar neutrinos are converted to active $\nu_{\tau}$ and 
atmospheric neutrino deficit is due to     
$\nu_{\mu} - \nu_s$ oscillations. 

The atmospheric neutrino problem gets resolved in terms of the 
$\nu_\m\rightarrow \nu_s$ oscillations also in the scenario
proposed in \cite{LS}. However
solar neutrinos 
are converted to the state which has 
a comparable admixture of the active and sterile components. 
These possibilities can be distinguished by studying  the neutral current 
interactions in the solar neutrinos (SK, SNO experiments) 
as well as NC induced events ($\pi^0$) and zenith angle 
distributions in the atmospheric neutrinos. 
In fact in the model \cite{LS} there is no explanation of the 
LSND result for the generic values of parameters. 

The hierarchy studied here was also proposed in \cite{hier} purely
phenomenologically. We have analysed this hierarchy in the context of
specific see-saw models with Grand Unification. The difference between
purely see-saw predictions and the present hierarchy can 
arise from the physics at the intermediate scale such as the PQ breaking. 
Also it is possible that all light scales are associated with 
GU scale and the scale of SUSY breaking.

The same pattern of mixings and masses as proposed here was shown to arise 
in a model \cite{jose} with purely radiatively generated neutrino masses.
The DM neutrino mass appears in one loop 
whereas the solar neutrino scale is generated in two loops. 
For certain choice of parameters the splitting between the two heavy
states can solve the atmospheric neutrino problem. 
It is not always easy to include grand unification into such models which
postulate an existence of new charged scalar 
bosons $k^{++}$,  $\eta^{+}$ and $h^{+}$ at the electroweak scale. 
This model can be distinguished by detection of the effects of these new
bosons.

A see-saw type of scheme with the mass hierarchy similar to ours
and based on the singular see-saw mechanism has been considered in
\cite{chun2} also. The right handed neutrinos are required to be very
light
($\sim \keV$) in \cite{chun2} unlike the standard GU scenario pursued
here. The eV mass scale for the neutrinos also does not come out
naturally.

In conclusion, the presence of a non-zero neutrino mass 
as hinted by the solar neutrino
observations may ultimately be pointing to the underlying grand unified
structure. We proposed a minimal extension of this basic structure
which also accommodates explanations for (i) the atmospheric neutrino
anomalies (ii) LSND results and (iii) provides hot dark matter in the
universe. The scenario suggested here can be tested in future solar
neutrino experiments as well as in atmospheric neutrino observations at
Superkamioka with increased statistics. The basic scenario proposed here
could arise from an underlying supersymmetric theory. Thus verification
of the scenario may indirectly probe supersymmetry as well as grand
unification.\\

%%%%%%%%%%%%%%%%%%%%%%%%%%%%%%%%%%%%%%%%%%%%%%%%%%%%%%%%%%%%%%%%%%%%%%%%%%

{\bf Acknowledgments} 

We are grateful  to G. Dvali and S. Rindani for useful discussions. 
%%%%%%%%%%%%%%%%%%%%%%%%%%%%%%%%%%%%%%%%%%%%%%%

\begin{figure}[t]
\epsfxsize 20cm
\epsfysize 20cm
\epsfbox[25 251 625 804]{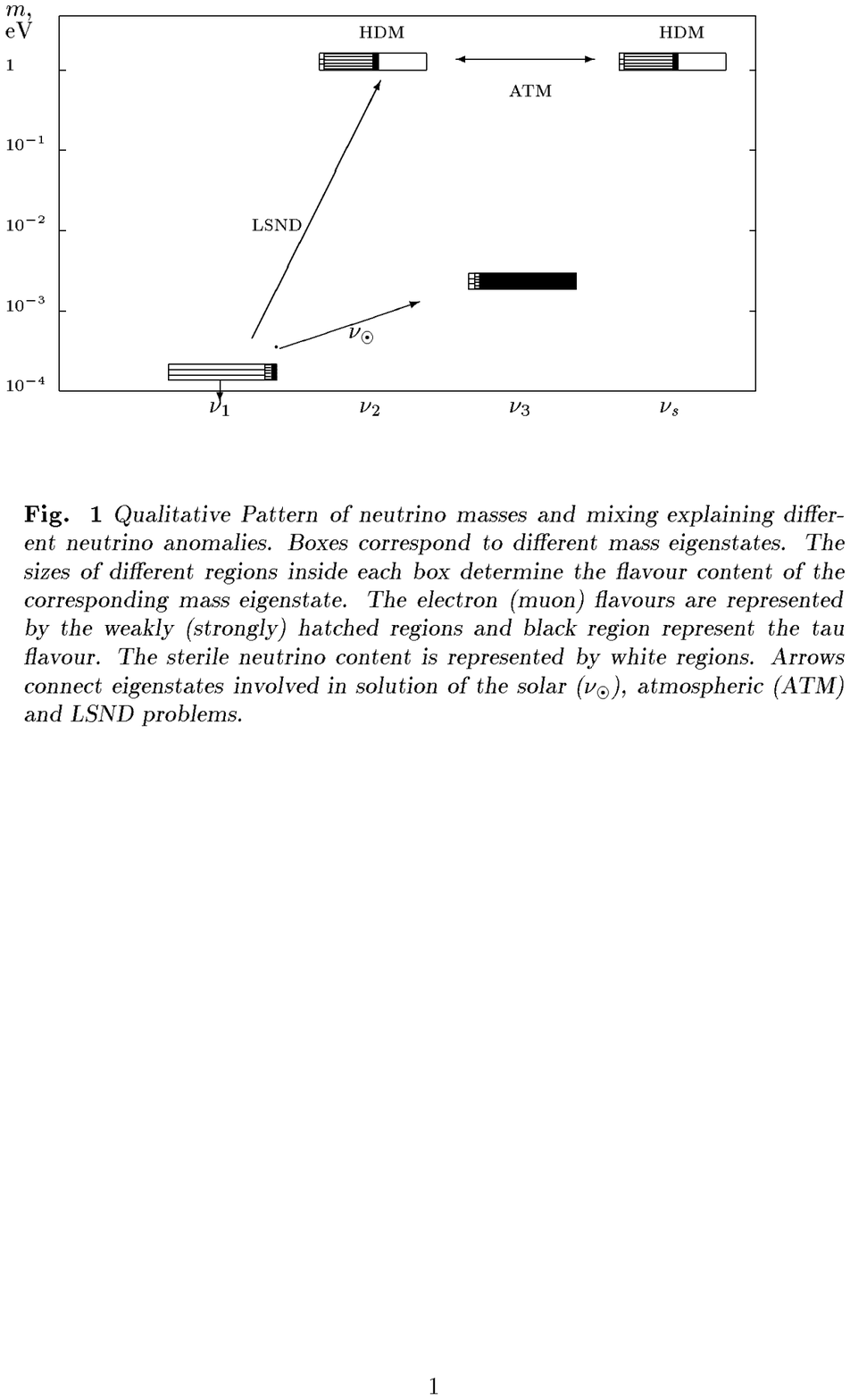}
\end{figure}
\end{document}